\documentclass[graybox]{svmult}

\usepackage{graphicx}
\usepackage{multicol}
\usepackage{newtxtext}
\usepackage{newtxmath}
\usepackage{url}
\newcommand{\beq}{\begin{equation}}
\newcommand{\eeq}{\end{equation}}
\newcommand{\beqa}{\begin{eqnarray}}
\newcommand{\eeqa}{\end{eqnarray}}
\newcommand{\ba}{\begin{array}}
\newcommand{\ea}{\end{array}}
\usepackage{color}

\makeindex            
\begin{document}
\title*{Attractive Multidimensional Solitons in Trapping Potentials}
\titlerunning{Attractive Multidimensional Solitons}
\author{
Fatkhulla Abdullaev
\and
Mario Salerno
}
\institute{
Fatkhulla Abdullaev
\at
 Physical-Technical Institute of Uzbekistan Academy of Sciences
\email{fatkhulla@yahoo.com}
\and
Mario Salerno
\at
Department of Physics "E.R. Caianiello", University of Salerno
\email{salerno@sa.infn.it}
}
\maketitle
\abstract{
This chapter reviews theoretical advances on the formation and stabilization of multidimensional solitons in nonlinear Schr\"odinger systems with attractive interactions, focusing on atomic Bose-Einstein condensates and nonlinear optics. While 1D solitons are generally stable, their 2D and 3D counterparts are prone to collapse. Several mechanisms have been proposed to mitigate this, including optical lattices, modulation of the nonlinearity via Feshbach resonance management, and Rabi coupling between hyperfine states. Other approaches involve competing nonlinearities and quantum corrections, such as Lee-Huang-Yang effects. Emphasis is placed on conditions enabling long-lived or fully stable solitons. Despite experimental feasibility, achieving robust stabilization remains challenging due to the intricate interplay of nonlinearities and external controls. The chapter surveys collapse dynamics, stabilization strategies, and soliton existence based on key theoretical contributions.
}
\section{Introduction}

Localized structures in the form of two- and three-dimensional solitons are central to nonlinear phenomena in atomic 
Bose-Einstein condensates (BECs) and nonlinear optics. While one-dimensional solitons are typically stable under 
attractive interactions, higher-dimensional ones are prone to collapse and instability~\cite{Kuznetsov-1986, 
Kagan-1998,Sulem-1999}. To address this, various stabilization strategies have been developed over the past two 
decades~\cite{Baizakov-2003, Malomed-2005, Saito-2007}.

Early approaches include the use of impurity potentials~\cite{Gaididei-2001} and optical lattices 
(OLs)~\cite{Morsh,Konotop}. It was shown that multidimensional (multi-D) OLs can arrest collapse and stabilize solitons 
both two- (2D) and three-dimensional (3D) cases~\cite{Baizakov-2003}. Interestingly, stabilization can also occur with 
reduced lattice dimensionality, such as 1D OLs in 2D~\cite{Baizakov-2004} and 2D OLs in 3D~\cite{Baizakov-2004, Mihalache-2004}. Radial lattices, inducing concentric modulations, also support stable soliton families~\cite{Kartashov-2004, 
Mihalache-2005, Baizakov-2006}.

Beyond fundamental solitons, vortex-solitons with quantized angular momentum have been studied 
in OLs~\cite{Kartashov-2011}. These are sensitive to azimuthal instabilities but can be stabilized by square OLs~\cite{Baizakov-2003, Mayteevarunyoo-2009}, where vorticity remains meaningful despite broken axial symmetry.

Stabilization can also be achieved through modulation of the nonlinearity. Feshbach resonance management (FRM) uses 
time-periodic magnetic fields to alternate between attractive and repulsive interactions, stabilizing 2D solitons 
under suitable parameters~\cite{Abdullaev-2003, Saito-2003}, 
and leading to  stable pattern formation in 2D BEC   \cite{Enns,Chin}. 

Spatially modulated nonlinearities, so-called nonlinear optical lattices (NOLs) \cite{KarMal}, induced via optically controlled Feshbach resonances~\cite{OFR1, OFR2} support stable 1D solitons, with more limited success in 2D~\cite{Kartashov-2011}. Combined OL–NOL systems allow stable solitons in 2D and 3D~\cite{daLuz-2010, Abdullaev-2012}, and recently, NOLs were shown to support self-trapped 2D dynamics~\cite{Ismailov-2024}.

Another approach, Rabi management, leverages radio-frequency coupling between atomic states of opposite scattering 
lengths~\cite{Saito-2007, Abdullaev-2008}. Under proper tuning, oscillations between these states yield a dynamic 
balance that supports stable 2D solitons.

Additional mechanisms include competing nonlinearities—such as cubic–quintic or saturable types—which can suppress 
collapse \cite{Akhmediev,Abdullaev-2001,Bulgac} and support both fundamental and vortex solitons in 2D and 3D~\cite{Malomed-2002, AG}. Spin–orbit coupling 
(SOC) in multicomponent BECs also supports 2D solitons and quantum droplets via the interplay of SOC-induced 
dispersion and interactions~\cite{Li-2017}. Quantum corrections like the Lee–Huang–Yang (LHY) term~\cite{LHY-57} 
further enrich stabilization scenarios, allowing for self-bound droplets in dipolar BEC and Bose–Bose 
mixtures~\cite{Petrov-2015,Luo-2021}.

Most of these solitons are experimentally realizable. Key tools, optical lattices, and magnetic or optical Feshbach 
tuning, are well established in ultracold atom experiments. However, the practical realization of robust 3D solitons 
remains experimentally challenging.

This chapter reviews key results—primarily theoretical—on collapse and soliton stabilization in multi-D nonlinear Schrödinger equations (NLSE) with attractive interactions. Section~2 addresses collapse in single- and two-component Gross-Pitaevskii equations (GPEs), including weak and strong collapse and modulational instability in 2D. Section~3 discusses stabilization mechanisms involving three-body interactions and quantum fluctuations, including 2D Townes solitons. Section~4 focuses on OL-supported multidimensional solitons, including radial and hybrid lattice configurations. Section 5 describes multi-D gap solitons in radially periodic potentials. Section~6 analyzes stabilization via dynamic management of system parameters, such as time-periodic nonlinearities and Rabi coupling. We conclude in Section~7 with a short summary and perspectives for future research.

\section{Multi-D Nonlinear Schr\"odinger Equations with Attractive Interactions}
In the context of dilute quantum gases, the NLSE appears in the form of the Gross-Pitaevskii equation (GPE), which governs the evolution of the macroscopic wavefunction of a BEC in the mean-field approximation~\cite{Pitaevskii-Stringari-2003}. Despite their different origins, the GPE and the NLS equation share the same mathematical structure, which enables the transfer of theoretical insights between these fields.

In this section, we discuss how attractive interactions in two and three dimensions lead to collapse and instability. We highlight experimental realizations of these effects in ultracold atomic BECs and distinguish between two types of collapse dynamics: {\it weak} and {\it strong} collapse (see below).

The equation for the time-dependent condensate wavefunction in the mean-field approximation—the Gross-Pitaevskii equation (GPE) is:
\beq
{\rm i} \frac{\partial \Psi}{\partial t} = 
-\frac{\hbar^2}{2m}\nabla^2\Psi + V({\bf r},t) \Psi + g
|\Psi|^2\Psi,
\label{3DGPE0}
\eeq
where $\nabla^2$ denotes the three-dimensional Laplacian, $V({\bf r},t)$ is a generic trap potential, $m$ is the atomic mass, and $g = \frac{4\pi \hbar^2}{m}a_s$ is the mean-field interaction strength, with $a_s$ being the $s$-wave scattering length. The wavefunction $\Psi \equiv \Psi({\bf r},t)$ is normalized to the total number of atoms.

Depending on the sign of $a_s$, the interaction is either repulsive ($a_s > 0$) or attractive ($a_s < 0$). In the absence of an external trap, Eq. \eqref{3DGPE0} reduces to the standard NLSE. This structural similarity implies that many nonlinear wave phenomena known in optics and hydrodynamics—such as modulational instability, soliton formation, and collapse—also manifest in the physics of Bose-Einstein condensates.

\subsection{Weak and strong collapse in 2D and 3D NLSE}

In the absence of an external trap, a BEC with attractive two-body interactions described by the NLSE undergoes wavefunction collapse, classified as "weak" in 2D and "strong" in 3D.

Weak collapse involves only a small fraction of the condensate norm and is typically localized, allowing most atoms to survive. It can lead to stable remnants such as solitons or fragmented wave packets. Experimental evidence includes soliton trains formed after controlled collapses~\cite{Hulet1, Everitt-2017}.

Strong collapse in 3D leads to a self-similar concentration of the condensate into an infinitesimal region, resulting in finite-time singularities in mean-field models. Though regularized in physical systems, it is often accompanied by rapid atom loss and explosive dynamics, as observed in Bose-Nova experiments~\cite{Cornish-2006}.

A foundational analysis by Kuznetsov et al.~\cite{Kuznetsov-1986} showed that collapse in the focusing cubic NLSE 
depends on both dimensionality and norm: in 2D it occurs only if the norm exceeds a critical value $N_{\rm cr}$ 
(weak collapse), whereas in 3D it can take place for any positive norm (strong collapse), with solutions becoming 
singular in finite time.

A qualitative understanding of collapse can be obtained by analyzing the system's total energy. In $D$ spatial dimensions, the total energy of the condensate is given by
\begin{equation}
E = \int d^D r \left[ \frac{\hbar^2}{2m} |\nabla \Psi|^2 + V({\bf r}, t) |\Psi|^2 + \frac{g}{2} |\Psi|^4 \right].
\end{equation}
Applying the scaling transformation
\begin{equation}
\Psi({\bf r}, t) = L^{-D/2} \phi\left(\frac{{\bf r}}{L}\right),
\end{equation}
yields a rescaled energy functional of the form:
\begin{equation}\label{eq:E}
E(L) = \frac{A}{L^2} - \frac{B}{L^D},
\end{equation}
where $A$ and $B$ are positive constants depending on the specific shape of $\phi$ and system parameters.

In 1D, the energy $E(L)$ exhibits a minimum, corresponding to a stable soliton and thereby excluding the possibility of collapse.  
In 2D, the behavior is marginal:
\begin{equation}
E(L) \sim \frac{N(N - N_{\rm cr})}{L^2},
\end{equation}
where $N_{\rm cr}$ is the critical norm for collapse. Variational analysis gives $N_{\rm cr} \approx 1.8$, while numerical simulations yield $N_{\rm cr} \approx 1.86$. In 3D, the energy has a local maximum at finite $L$, indicating an inherent instability toward collapse.

External potentials, like harmonic traps or narrow local inhomogeneities \cite{Gaididei-2001} can regularize collapse and stabilize multidimensional matter waves. For example, a harmonic trap $V({\bf r}) = \alpha r^2$ introduces a stabilizing term $\sim C L^2$ in Eq.~\eqref{eq:E}, leading to a minimum of $E(L)$ and the existence of a metastable state in 3D. The critical number of atoms in a harmonic trap is given by~\cite{Kagan-1998}:
\begin{equation}
\label{eq:k}
N_{\rm cr}=k\frac{a_{\rm ho}}{|a_s|},
\end{equation}
where $a_{\rm ho} = \sqrt{\hbar / (m \omega_{\rm ho})}$ is the harmonic oscillator length, and $k$ is a dimensionless constant dependent on the geometry and interaction.

Several mechanisms have been proposed to arrest collapse in BECs, including optical lattices~\cite{Baizakov-2003}, 
three-body interactions~\cite{Abdullaev-2001}, scattering length modulation via Feshbach resonances~\cite{Saito-2003, 
Abdullaev-2003, Montesinos}, quantum fluctuations~\cite{Petrov-2015, AP}, 
Rabi coupling~\cite{Saito-2007, Abdullaev-2008}, and spin-orbit coupling~\cite{MalSher}. These mechanisms are discussed in detail in the following sections.

Collapse has also been experimentally studied in attractive BECs~\cite{Coll1, Coll2, Donley-2001}. In~\cite{Coll2}, 
the critical atom number $N_{\rm cr}$ yielded a value $k = 0.461 \pm 0.012$, slightly below the theoretical $k = 
0.574$. Collapse occurred in bursts, ejecting atoms from the condensate. Remnants observed in~\cite{Altin-2011} often 
appeared fragmented or soliton-like, exhibiting repulsive interactions consistent with a post-collapse soliton gas. 
Collapse times, modeled via Gross-Pitaevskii simulations with three-body loss terms $K_3 |\psi|^4 \psi$, decreased 
with increasing $|a_s|$ as expected. Post-collapse dynamics featured oscillatory behavior and Bose-Nova bursts.

\subsection{Modulational instability. Nonlinear stage of evolution.}
Modulational instability usually refers to the instability of nonlinear plane waves with respect to weak long-wavelength modulations whose wavenumbers lie below certain critical values. This phenomenon is of great interest because it can lead to the generation of different types of solitons.

The process of modulational instability (MI) in the two-dimensional GP equation can be analyzed by considering the perturbed nonlinear plane-wave solution
\begin{equation}
(A + \psi_1(r, t)) e^{i g A^2 t},
\end{equation}
which leads to the expression for the MI gain
\begin{equation} 
G = k(2 g A^2 - k^2)^{1/2},\;\;\;  k^2 =k_x^2+k_y^2.
\end{equation}

The maximum of the gain occurs at $ k=k_m=(g A^2)^{1/2}$.
In the case of a cigar-shaped trap, the quasi-1D non-polynomial GPE can be used to analyze the process of matter-wave soliton generation by MI~\cite{Everitt-2017}. The maximum of the MI gain occurs when the wavenumber of the modulations is $k_m$
and the healing length is given by $\xi \sim 1/k_m$. The average distance between peaks (solitons) is $2\pi \xi$, so the number of generated solitons can be estimated as~\cite{Hulet2,Everitt-2017}
\begin{equation}
N_s \sim \frac{L_x}{2\pi \xi}.
\end{equation}
If modulations are seeded by noise, quantum fluctuations, or similar mechanisms, the growth of components starts from the center of the initial broad flat distribution. The generated solitons then have phase differences $ \pi/2 \leq \Delta\phi \leq \pi$, corresponding to repulsive interactions between them. This mechanism was investigated in~\cite{Salasnich-2003}.
If instability occurs due to the growth of fringes from the ends of the distribution, where interference and nonlinearity are responsible for the process,  solitons are generated from the ends of the distribution, and the center becomes depleted. This mechanism was studied in~\cite{Kamch,Clark}.

An experiment demonstrating the formation of solitons during BEC collapse in a 3D trap was done in~\cite{Cornish-2006}. It was shown that  a trapped BEC with attractive interactions collapses if the number of atoms exceeds the critical value $N_{cr}$ is given by Eq.(\ref{eq:k}).
The experiment was done with $^{85}$Rb and an initial number $N=15,000$ atoms. It was observed that after the initial injection of atoms, the subsequent collapse left a stable remnant with $N_{remn} \sim 10 N_{cr}$, which remained stable for several seconds. Investigations showed that this state corresponded to multiple solitons with repulsive interactions between them, and that the number of atoms in each soliton was below $N_{cr}$.

\subsection{Collapse in two-component BEC systems}

Two-component ($\psi_{1,2}$) untrapped BECs in $D$ dimensions are commonly described by the coupled Gross-Pitaevskii equations:
\begin{eqnarray}
i\hbar\psi_{1,t} &=& -\frac{\hbar^2}{2m_1}\nabla^2 \psi_1 + (\tilde{g}_{11} |\psi_1|^2 + \tilde{g}_{12}|\psi_2|^2)\psi_1 -\hbar\Omega\psi_2,\nonumber \\
i\hbar\psi_{2,t} &=& -\frac{\hbar^2}{2m_2}\nabla^2\psi_2 + (\tilde{g}_{22} |\psi_2|^2 + \tilde{g}_{21}|\psi_1|^2)\psi_2 -\hbar\Omega\psi_1.
\end{eqnarray}

Here, $\tilde{g}_{ij} = 4\pi\hbar^2 a_{ij}/m$ for $D=3$ and $2\sqrt{2\pi}\hbar^2 a_{ij}/(m l_z)$ for $D=2$, where $a_{ij}$ are the intra- and inter- species scattering lengths, and $l_z = \sqrt{\hbar/(m \omega_z)}$ is the transverse harmonic oscillator length. $\Omega$ is the Rabi frequency.

Assuming equal masses $m_1 = m_2 = m$ and introducing dimensionless variables:
\begin{equation}
(x,y) = (\tilde{x},\tilde{y})/l_z, \quad t = \tilde{t}\omega_z, \quad
u = 4\pi l_z a_{ij}^{(0)}\psi_1, \quad v = 4\pi l_z a_{ij}^{(0)}\psi_2, \quad
g_{ij} = 4\pi l_z a_{ij}/a_{ij}^{(0)},
\label{eq:dim_units}
\end{equation}
where $a_{ij}^{(0)}$ is a reference scattering length, the above equations become:
\begin{eqnarray}\label{eq:original_model}
i\frac{\partial u}{\partial t}+ \frac{1}{2}\nabla^2 u - (g_{11}| u|^2 +g_{12}| v|^2)u + \Omega v &=& 0, \nonumber\\
i\frac{\partial v}{\partial t}+ \frac{1}{2}\nabla^2 v - (g_{22}| v|^2+ g_{21}| u|^2)v + \Omega u &=& 0.
\end{eqnarray}
In nonlinear optics, these equations also describe spatial solitons, where $t$ plays the role of the propagation coordinate, and $u$, $v$ represent two mutually incoherent light beams.

For $\Omega = 0$, the Hamiltonian of the system is:
\begin{equation}\label{eq:Hamiltonian}
H = \int_R \left( \frac{1}{2}|\nabla u|^2 + \frac{1}{2}|\nabla v|^2 + \frac{1}{2}g_{11}|u|^4 + \frac{1}{2}g_{22}|v|^4 
+ g_{12}|u|^2|v|^2 \right) d^D x.
\end{equation}

In 2D, with $\nabla^2 = \partial_x^2 + \partial_y^2$ and $g_{12} = g_{21}$, the virial theorem yields a blow-up 
criterion~\cite{Vlasov-1971,McKinstrie-1988,Bang-1998,Vekslerchik-2009}:
\begin{equation}
\frac{d^2 I}{dt^2} = \frac{4H}{N},
\end{equation}
where $I=\int \rho^2 (|u|^2 + |v|^2)d^2x,\rho^2=x^2 + y^2$, and indicating collapse when $H(0) < 0$.

For a Gaussian ansatz,
\[
u = \left( \frac{N_1}{\pi w^2} \right)^{1/2} e^{-{\rho^2}/{2w^2}}, \quad  v = \left( \frac{N_2}{\pi w^2} \right)^{1/2} e^{-{\rho^2}/{2w^2}},
\]
the initial energy is:
\begin{equation}
H(0) = \frac{1}{2w^2} \left( N - \frac{g_{ik}}{2\pi} N_i N_k \right),
\end{equation}
from which the critical norm $N_{cr}$ can be determined.

 The collapse condition for a two-component condensate in 2D reads~\cite{Vekslerchik-2009}:
\begin{equation}
g_{12} \le -\sqrt{g_{11} g_{22}},
\end{equation}
where $g_{11}$ and $g_{22}$ are the intra-species interaction coefficients.  Notably, even when both intra-species interactions are repulsive ($g_{11}, g_{22} > 0$), collapse can still occur if the inter-species attraction is sufficiently strong to satisfy the above inequality. This condition is reminiscent of the miscibility criterion in two-component BECs \cite{Baizakov-2025}, which also arises from the competition between intra- and inter-species interactions, although here it governs the onset of collapse rather than phase separation.

\section{Collapse Stabilization.}
In this section we discuss three-body nonlinear interactions and quantum fluctuations as two possible mechanisms to stabilize collapse. In particular, we show that both three-body interactions and beyond-mean-field quantum corrections, like the Lee-Huang-Yang (LHY) term, introduce effective repulsion that can stabilize collapsing states. An example of weak collapse stabilization will be given below for the case of 2D Townes solitons.

\subsection{Collapse stabilization via three-body interactions}
The inclusion of repulsive three-body interactions into the GPE can stabilize the collapse of an attractive Bose-Einstein condensate (BEC). In this case, the total energy density of the system can be approximated as
\begin{equation}
E(n) \sim -\alpha n^2 + \beta n^3,
\end{equation}
where $n = |\psi|^2$  is the condensate density, and the coefficients $\alpha,\beta$  are proportional to the strengths of the two-body attractive and three-body repulsive interactions, respectively. The competition between these two contributions leads to an energy minimum at a finite equilibrium density. This equilibrium density, corresponding to a self-bound state, is given by
\begin{equation}
n_c \sim \frac{2\alpha}{3\beta}.
\end{equation}
Such a balance results in the formation of localized, self-trapped condensates, referred to as quantum droplets, that are stabilized against collapse by the effective repulsion from the three-body interactions~\cite{Abdullaev-2001,Akhmediev,Bulgac}. These structures exhibit liquid-like behavior and represent a novel phase of matter in ultracold atomic systems (see next subsection).

\subsection{Stabilization via quantum fluctuations}

Another important mechanism for the dynamical stabilization of collapse arises from quantum fluctuations~\cite{Petrov-2015,AP}. These beyond-mean-field effects are described by the Lee–Huang–Yang (LHY) correction to the energy density.
For a single-component condensate in 3D, the LHY term reads:
\begin{equation}
\mathcal{E}_{\mathrm{LHY}}^{(3D)} = \frac{8}{15\pi^2} 
\left( \frac{m}{\hbar^2} \right)^{3/2} (g n)^{5/2},
\end{equation}
where \( g = 4\pi \hbar^2 a / m \) is the coupling constant, \( a \) the scattering length, and \( n=|\psi|^2 \) the density.

In 2D, quantum fluctuations for a binary mixture with equal 
intra-species scattering lengths, \(a_{11}=a_{22}=a\), give a 
logarithmic contribution~\cite{Petrov-2015}:
\begin{equation}
\mathcal{E}_{\mathrm{LHY}}^{(2D)} = 
\frac{8\pi n^2}{\ln^2(a_{12}/a)} 
\ln\!\left( \frac{n}{e n_0}\right), 
\qquad 
n_0 = \frac{e^{-2\gamma-3/2}}{8\pi}\frac{\ln(a_{12}/a)}{a a_{12}},
\end{equation}
where $a_{12}$ is the inter-species scattering length and $\gamma$ is the Euler constant.

For general binary mixtures in 3D, the LHY correction takes the form:
\begin{equation}
\mathcal{E}_{\mathrm{LHY}}^{\mathrm{mix}} = 
\frac{8}{15\pi^2} \left( \frac{m}{\hbar^2} \right)^{3/2} 
\left( g_{11} n_1 + g_{22} n_2 \right)^{5/2} 
f\!\left( \frac{g_{12}^2}{g_{11}g_{22}}, \frac{n_1}{n_2} \right),
\end{equation}
where \( g_{ij} = 4\pi \hbar^2 a_{ij}/m \) are the interaction strengths, and \( f \) is a dimensionless function depending on the density ratio and interaction parameters.

Thus, depending on dimensionality and the number of components, the LHY term provides an effective repulsion that can counteract mean-field attraction and prevent collapse, giving rise to stable self-bound quantum droplets.

In practice, the full expression for the LHY correction in mixtures is complicated. Therefore, in many theoretical works a simplified scalar model is employed, where the  beyond-mean-field term is approximated as:
\begin{equation}
\mathcal{E}_{\mathrm{BMF}} = \gamma_{\mathrm{QF}} n^{5/2},
\end{equation}
with \( \gamma_{\mathrm{QF}} \) an effective coefficient 
encoding the strength of quantum fluctuations 
for a near-balanced mixture. In this reduced description, the total energy density is written as:
\begin{equation}
\mathcal{E}(n) = -\alpha n^2 + \gamma_{\mathrm{QF}} n^{5/2},
\end{equation}
where the effective attraction parameter \( \alpha \sim (a - a_{12}) \) 
quantifies the small imbalance between intra- and inter-species interactions. Minimizing this energy functional with respect to $n$ yields the equilibrium density of a quantum droplet as:
\begin{equation}
n_c = \left( \frac{4\alpha}{5\gamma_{\mathrm{QF}}} \right)^2.
\end{equation}

From this picture the physical mechanism of stabilization becomes clear: weak mean-field attraction is counterbalanced by the repulsive LHY term, interpreted as the zero-point energy of the Bogoliubov vacuum. 
Unlike standard bright solitons, which rely on integrability or external confinement, 
quantum droplets are self-bound, possess finite compressibility, and display liquid-like 
properties. Their existence has been confirmed experimentally both in dipolar condensates 
and in Bose–Bose mixtures~\cite{expQD1,expQD2}, marking a milestone in the study of stabilization mechanisms in ultracold quantum gases.

\subsection{Townes solitons stabilization via quantum fluctuations}
In two-dimensional nonlinear systems with attractive cubic interactions, such as the 2D nonlinear Schrödinger equation (NLSE), a special class of self-trapped, localized solutions—known as Townes solitons(TS) exists~\cite{Chiao-1964, Weinstein-1983}. These radially symmetric wave packets represent a critical threshold between dispersion and collapse and are solutions to the dimensionless stationary NLSE:
\begin{equation}
i\phi_t + \nabla^2 \phi + |\phi|^2 \phi = 0,
\end{equation}
where stationary solutions take the form $\phi(r,t) = e^{it} R(r)$, and $R(r)$ satisfies:
\begin{equation}
- R(r) + \nabla^2 R(r) + R^3(r) = 0,
\label{TS}
\end{equation}
with boundary conditions $R(\infty) = 0$ and $dR/dr|_{r=0} = 0$. The Townes soliton is the ground-state solution $R_T(r)$ of Eq.~(\ref{TS}) and exists only at a critical value of the norm:
\begin{equation}
N = \frac{1}{2\pi} \int_0^{\infty} d^2r\, R_T^2(r) \equiv N_{\mathrm{cr}} \approx 1.862.
\end{equation}
This \textit{isolated existence} implies marginal stability: for $N < N_{\mathrm{cr}}$, the wave packet disperses, while for $N > N_{\mathrm{cr}}$, the system undergoes weak collapse due to the scale invariance of the 2D cubic NLSE. The virial relation $\frac{d^2}{dt^2} \langle (x^2+y^2) \rangle = 8 H(0)$ ~\cite{Vlasov-1971}, with $H(0)$ the initial energy, implies collapse for $H(0)< 0$, and spreading for $H(0)>0$.

Townes solitons have been observed in ultracold atomic gases in two main experimental settings~\cite{Cheng, BBH}, both not involving quantum fluctuations. 

The possibility of stabilization the Townes soliton via quantum fluctuations has recently been demonstrated for the 
two-dimensional system of $N$ bosons coupled with attractive interactions. Quantum fluctuations allow the formation of 
a stable two-dimensional \textit{self-bound  state of N-bosons}~\cite{Petrov24}. The total energy in this case
becomes:
\begin{equation}\label{eq:n-body}
E(R) = \frac{(g - g_c) N^2}{2\pi N_{cr} R^2} - \frac{N_{cr}}{4 R^2} \ln\left( \frac{\xi R}{h} \right),
\end{equation}
where $g_c$ is the critical coupling at which quantum and mean-field terms balance, $\xi$ is a numerical constant,
and  $h$ is a short-distance cutoff. The resulting energy minimum at finite radius $R$ indicates a stable 
configuration:
\begin{equation}
R = \frac{1}{|B_2|} \exp\left( -\frac{2N}{N_{cr}} + \frac{1}{2} + \ln(4\sqrt{2}) \right),
\end{equation}
where $B_2 < 0$ is the dimer (N=2) energy. The energy in the minimum point is:
\begin{equation}
B_N = E(R_N) = \frac{N_{cr}}{8 R_N^2} = B_2 \exp\left( \frac{4N}{N_{cr}} - 1 - 8 \ln 2 + \ln(N_{cr} \xi^2) \right).
\end{equation}
Note that the energy of N bosons is scaled via the dimer energy as $B_N \sim B_2\exp(4N/N_{cr}).$
The frequency of breathing oscillations of the Townes soliton induced by solely quantum fluctuations is proportional
to $\omega_B \sim |B_N|/\sqrt{N}$ for large $N$.

This result also follows from the generalized GP equation, when the logarithmic density dependence of the coupling constant in the quantum-fluctuation term is taken into account~\cite{Sukhorowski}. The application of modulation theory~\cite{Fibich-1999} to the Townes soliton perturbed by quantum fluctuations then leads to Eq.~(\ref{eq:n-body}). This mechanism allows the realization of robust Townes solitons beyond the marginally stable regime, providing a controlled route to collapse prevention in 2D quantum gases with attractive interactions.
\section{Multi-D Solitons in Different Periodic Potentials}
In this section, we demonstrate that periodic optical-lattice potentials can stabilize both 2D and 3D solitons. These 
results are relevant to BECs in OLs and to nonlinear optics, where they describe light beams propagating in bulk 
self-focusing Kerr media with transverse refractive-index modulation (photonic crystals). In the optical 3D case, solitons are usually treated in 2D transverse geometries with the third dimension serving as the evolution variable.

\begin{figure}[tbp]
\centerline{
\includegraphics[width=12.cm,clip]{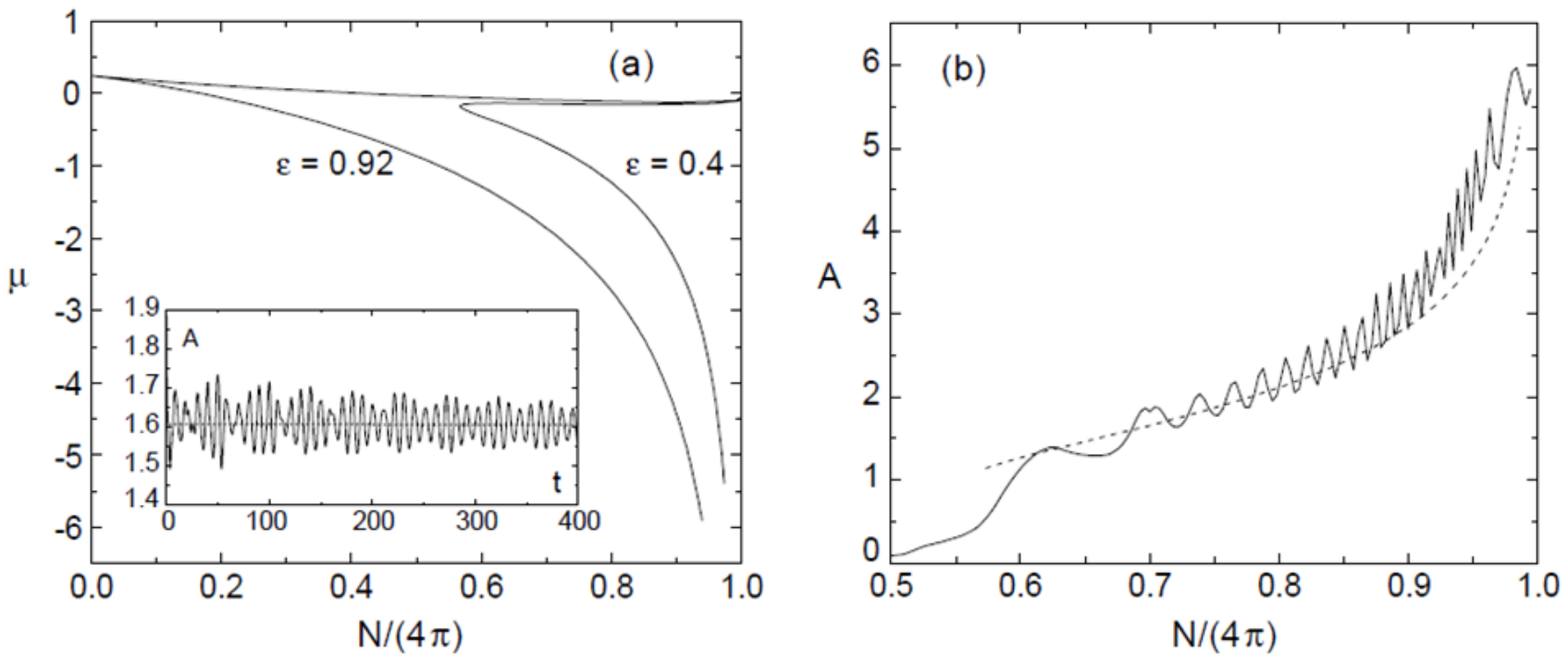}}
\vskip -1.5cm
\caption{(a) Numerical solution to Eqs.~(\ref{2Dvar}) for $\varepsilon =0.4$ and $\protect\varepsilon =\protect
\varepsilon_{0}=0.92$ (the latter value is the one at which $N_{\mathrm{thr}}$ vanishes). Inset displays evolution of 
the amplitude of a directly simulated solution initiated by the variational ansatz~(\ref{ansatz}) with $A$ and $a$ 
taken as solutions of Eqs. (\ref{2Dvar}), in the case of $\protect\varepsilon =0.92$, $N=2\protect\pi $, $a=1.3$ (the 
VK-stable branch). (b) Dependence of the amplitude $A$ of established 2D solitons vs. the initial norm $N$, as
obtained from direct simulations of Eq. (\ref{gpe}) starting with the configuration predicted by VA. The undulations 
in the dependence is a result of radiation loss in the course of the soliton formation. The dashed curve is the
dependence $A(N)$ as given by VA. The figure is reproduced from Ref.~\cite{Baizakov-2003}.}
\label{fig1}
\end{figure}

\subsection{Stable multidimensional solitons in optical lattices}
The model is based on the rescaled Gross-Pitaevskii equation \cite{ReviewDalfovo} for the wavefunction $\psi(\mathbf{r},t)$:
\begin{equation}
i \psi_t + \left[ \nabla^2 + V(\mathbf{r}) + |\psi|^2 \right] \psi = 0, \label{gpe}
\end{equation}
which also represents the NLSE equation in optics (with $t$ as the propagation coordinate). The external potential
\begin{equation}
V(\mathbf{r}) = \varepsilon [\cos(2x) + \cos(2y) + \cos(2z)],\quad \mathbf{r} = (x, y, z), \label{potential}
\end{equation}
model a cubic optical lattice (OL) with period $\pi$ (the trap is omitted in optical models). The only dynamical invariant is the norm, $N = \int |u(\mathbf{r})|^2 d\mathbf{r}$, representing
the atom number (BEC) or beam power (optics). Approximate analytical soliton solutions can be obtained by means of  the variational approximation (VA) \cite{VA-Malomed}, originally developed in optics and successfully extended to BECs, including OL settings \cite{Perez-Garcia-2000, Wadati-2001,Parola-2002}. For this we look for stationary solitons of the form $\psi(\mathbf{r},t)=\phi(\mathbf{r})\exp\left(-i\mu t\right)$ with $\phi(\mathbf{r})$ given by the following  Gaussian ansatz 
\begin{equation}
\phi(\mathbf{r}) = A \exp\left(- \frac{a r^2}{2} \right), \label{ansatz}
\end{equation}
with parameters $A$, $a > 0$ and $\mu < 0$ (we center the soliton at the OL minimum $\mathbf{r} = 0$, assuming $\varepsilon > 0$). In BECs, $\mu$ is the chemical potential; in optics, $-\mu$ is the propagation constant. Substituting the ansatz into the Lagrangian of the GPE (\ref{gpe}), and performing the spatial integration, yields an effective Lagrangian $L_{eff}\equiv L_{eff}(a,A,\mu)$ whose stationary conditions are: $\partial L_{eff}/\partial A = \partial L_{eff}/\partial a = 0$.

For 2D solitons, these conditions give a relation between the width $a$ and norm $N = \pi A^2/a$ and an expression for the chemical potential $\mu$:
\begin{equation}
N = 4\pi\left(1 - 2\varepsilon a^{-2} e^{-1/a} \right), \quad
\mu = 2\varepsilon (2a^{-1} - 1) e^{-1/a} - a.
\label{2Dvar}
\end{equation}
In the absence of a lattice ($\varepsilon = 0$), this recovers the known critical norm $N_\mathrm{cr} = 4\pi$ for unstable 2D solitons. With $\varepsilon \neq 0$, a minimum (threshold) norm appears:
\begin{equation}
N_\mathrm{thr} = 4\pi \left(1 - 8\varepsilon e^{-2}\right),
\label{thr}
\end{equation}
which is positive only for $\varepsilon < \varepsilon_0=e^2/8 \approx 0.92$.
Eq.~(\ref{2Dvar}) implies that the norm $N$ cannot exceed the critical value $N_{cr}=4\pi$, so VA predicts fundamental 2D solitons in the interval $N_{thr}<N<N_{cr}$ (with $N_{thr}\equiv 0$ for $\varepsilon >\varepsilon_0$). Stability can also be assessed using the Vakhitov-Kolokolov (VK) criterion~\cite{VK}, which requires $d\mu/dN<0$, with $\mu\equiv \mu(N)$ obtained from Eqs.~(\ref{2Dvar}).
%
\begin{figure}[tbp]
\centerline{
\includegraphics[width=12 cm]{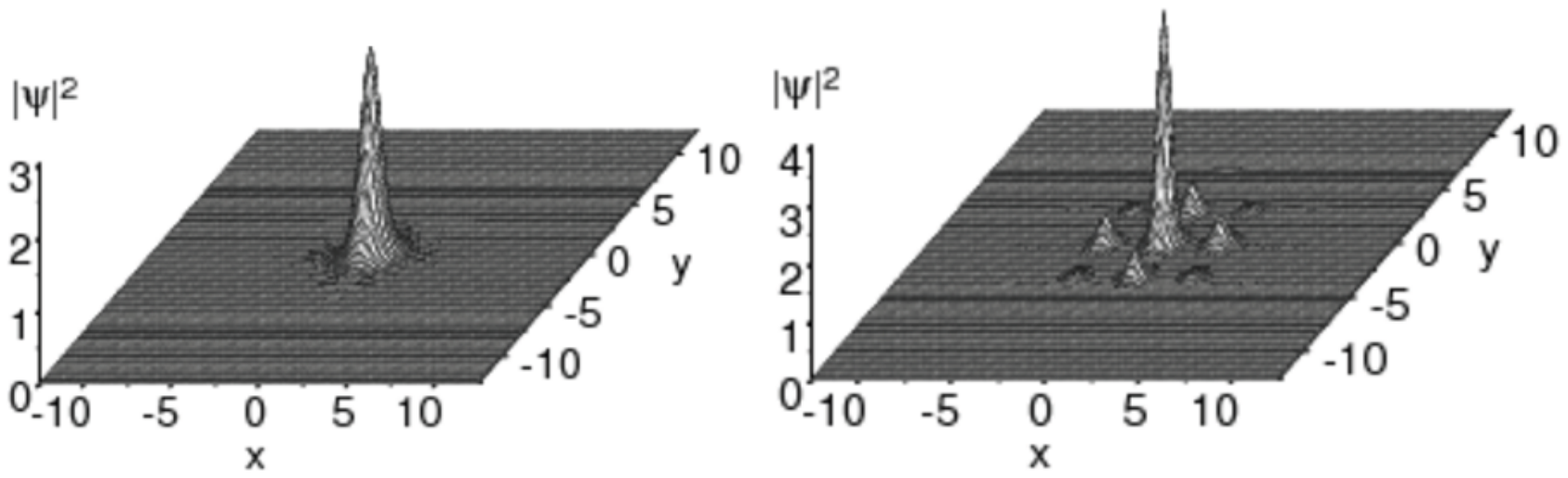}
}
\vskip -2cm
\caption{Left Panel. An established single-cell 2D soliton in a moderately strong
lattice potential, found from direct simulations of Eq. (\ref{gpe}) with
$\varepsilon =0.92$. The initial configuration was taken as per the
VA-predicted stable soliton, i.e., Eqs. (\ref{ansatz}) and (\ref{2Dvar})
were used, with $a=1.3$, the corresponding norm being $N=2\protect\pi $. Right panel.
Examples of 3D solitons formed in a strong OL.
The $z=0$ cross-section is shown for $\protect\varepsilon =10$ and $N=10$. 
Figure extracted from Ref.~\cite{Baizakov-2003}.
}
\label{fig2}
\end{figure}
Numerical solutions of Eqs. (\ref{2Dvar})  are shown in Fig.~\ref{fig1}a for two different values of
$\varepsilon$. Analytical results near the critical point $1- N/4\pi \rightarrow +0$ also give dual
solutions for $\mu$:
\beq
\mu_1(N) \approx -\left( \sqrt{\frac{2\varepsilon}{1 - N/4\pi}} + 2\varepsilon \right), \quad
\mu_2(N) \approx \left[ \ln(1 - N/4\pi) - \ln(2\varepsilon) \right]^{-1},
\label{mu}
\eeq
but only $\mu_1(N)$ satisfies the VK criterion, indicating stability. 
Direct simulations (see inset to Fig.~\ref{fig1}a and Fig.~\ref{fig2}) confirm that solitons along the VK-stable branch are dynamically stable and closely follow the VA predictions. The appearance of a finite $\mu$-band for stable solitons in the range $N_\mathrm{thr} < N < N_\mathrm{cr}$ stems from the presence of the periodic lattice potential. In the limit $\varepsilon \to 0$, this band shrinks to a point, recovering the standard 2D NLSE behavior. This is analogous to the existence of stable solitons with repulsive interactions in OL-supported models~\cite{Baizakov-2002}.

These VA predictions are corroborated by direct simulations of Eq.~(\ref{gpe}). Specifically, VA-based initial conditions evolve into stable solitons in both 2D and 3D, as illustrated in Fig.~\ref{fig2}. In contrast, inputs with norm below the VA threshold (for $\varepsilon < \varepsilon_0$), or significantly deviating from the ansatz, decay into dispersive waves. Such decay is shown in Fig.~\ref{fig1}b, near $N/(4\pi) \approx 0.62$ for $\varepsilon=0.4$, close to the VA threshold $N_{\min}/(4\pi) \approx 0.58$. Notably, the lattice prevents both decay and collapse: Fig.~\ref{fig1}b shows stable solitons even slightly above the critical norm $N_{\mathrm{cr}}/(4\pi) = 0.93$.
%
\begin{figure}[tbp]
\centerline{
\includegraphics[width=14cm]{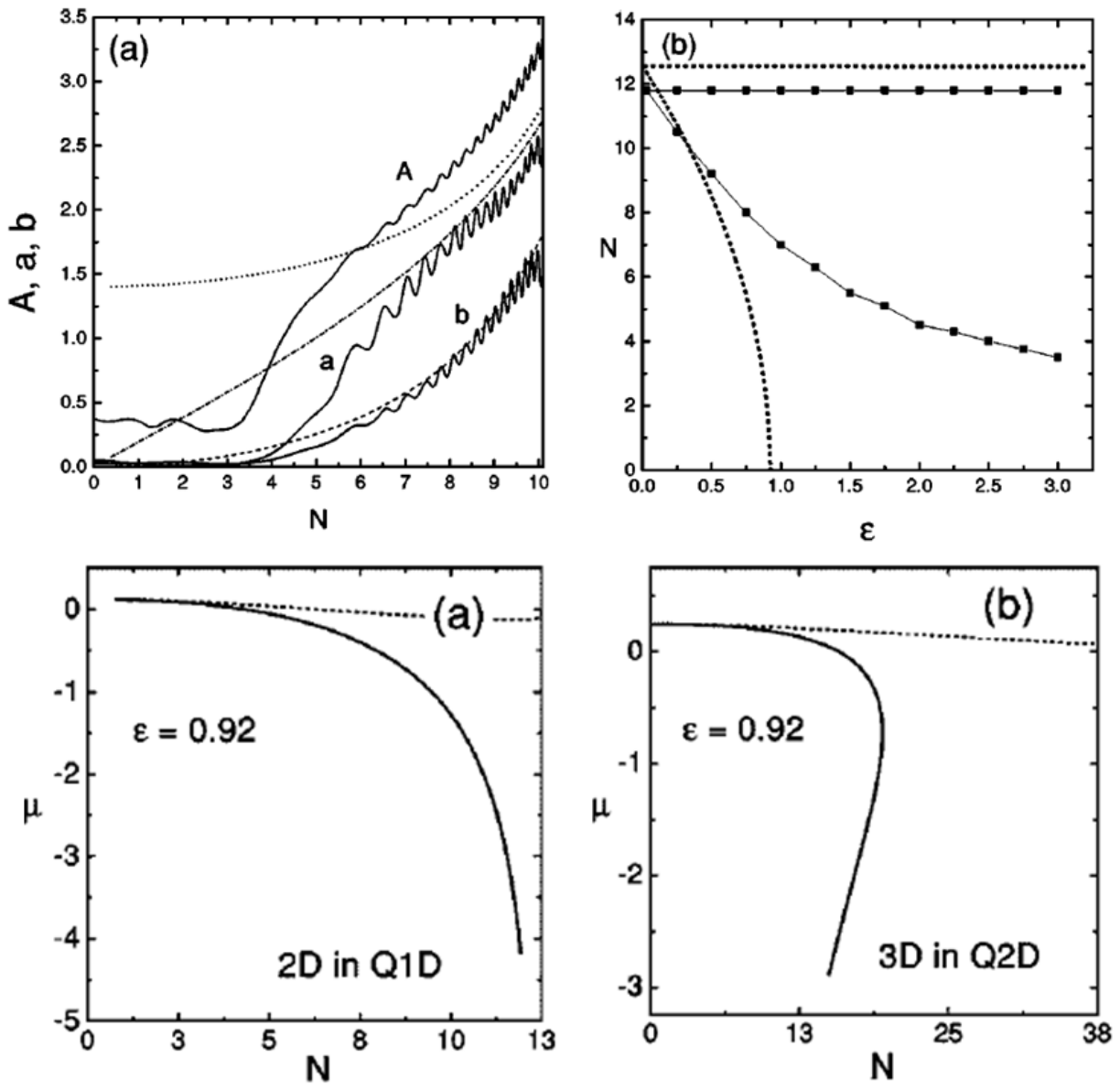}
}
\caption{ {\bf Top panels}.  {\bf (a)} : Variational parameters $A, a, b$ versus $N$ of stable 2D solitons in the quasi-1D potential with $\epsilon=2$, as found from numerical simulations 
of the Gross-Pitaevskii (solid lines) and as predicted by the VA for amplitude A (dotted line) and inverse squared 
widths, a and b, (dot dashed and dashed lines, respectively).  {\bf (b)}: The numerically found (connected squares) and VA
predicted (dashed lines) existence limits for stable 2D solitons in the quasi-1D potential.  {\bf Bottom panels} . The 
$\mu(N)$ dependence for:  {\bf (a)}: 2D solitons, and  {\bf (b)}:  3D solitons, in quasi-1D and quasi-2D potentials, respectively. In each panel, the solid and dashed curves show two different branches of the solution family. Figure extracted from 
Ref.~\cite{Baizakov-2004}.}
\label{fig3}
\end{figure}

\subsection{Multi-D solitons in low-dimensional potentials}
In this section we focus on multi-dimensional solitons in lower-dimensional periodic potentials,
specifically, quasi-1D (Q1D) and quasi-2D (Q2D) lattices in 2D and 3D settings. In optics, the 2D model
describes beam propagation in layered media with a transversely modulated refractive index, extending
earlier 1D multichannel systems. In BECs, the 2D and 3D cases correspond to condensates loaded into Q1D or
Q2D optical lattices. These configurations are relevant for two main reasons: (i) low-dimensional lattices are easier to
implement experimentally in both optics and BECs, and (ii) solitons can move freely along unconfined
directions, enabling studies of their collisions and bound states. Such interactions can give rise to interesting phenomena such as matter exchange, collapse and interference effects~\cite{Baizakov-2004}.

The variational analysis of the previous section, applied to the lower-dimensional periodic potentials of the form
\beq
V(x)= \varepsilon \cos(2x),\;\;\; V(x,y)= \varepsilon (\cos(2x)+\cos(2y))
\eeq
with stationary ansatze for the soliton of the form
\beq
\psi_{2D}= A e^{-\frac 12 (a x^2 + b y^2)}, \;\;\; \psi_{3D}= A e^{-\frac 12 (a (x^2+ y^2) + b y^2)},
\eeq
for the corresponding 2D and 3D GPE cases, respectively, show the existence of stable  attractive solitons. This can
be seen from Fig.~\ref{fig3} where  soliton parameters $A$, $a$, and $b$ in 2D, obtained from VA and from direct GPE 
simulations, are compared. We see from the top (a) panel that a good agreement is observed for $N \geq 5$; below this 
value (for $\varepsilon=2.0$), VA-predicted inputs evolve into delocalized states with amplitude decay and width 
growth. Also the $\mu(N)$ dependence depicted in the other panels of Fig.~\ref{fig3} make it possible to predict the 
solitons' stability on the basis of the VK criterion. These solitons existing within finite norm ranges predicted by 
the variational approximation (VA) are confirmed by simulations of the GPE (see~\cite{Baizakov-2004} for details). 

\subsection{Delocalization transitions of multi-D solitons}
In the previous subsection we have seen that the norm of multi-D solitons in periodic potentials is bounded not 
only from above by the onset of collapse, but also from below due to the delocalizing transition phenomenon
\cite{Kalosakas-2002,BS-2004}. The  
physical mechanism underlying the delocalizing transition can be ascribed to the extinction of bound states in the 
effective (soliton density) potential of  the underlying Schr\"odinger  equation. Since in the 1D case, any potential 
support at least one bound state, even for infinitesimal well depths, the above interpretation automatically implies 
that in 1D solitons cannot undergo delocalizing transitions. It is indeed well known that in 1D case by decreasing the 
strength of the OL or the coefficient of the nonlinearity, the soliton becomes more and more extended, and in the 
limit of zero nonlinearity, it reduces to a Bloch state, but always recovers its original shape when the parameters are 
reversed~\cite{Kalosakas-2002,BS-2004}. 
On the contrary, in  2D and 3D cases, there is always a critical values of system parameters below which the 
recovering of the soliton becomes impossible, i.e., the soliton irreversibly disintegrates into extended states. 
This can be understood as the absence of  bound states in the multi-D soliton effective potential. Indeed, by 
approximating the soliton effective potential with a suitable analytical potential, and adopting a variational ansatz
description for the soliton, one can show  that the bound-state interpretation leads to predictions of the 
delocalizing transition which are in good agreement with numerical integrations of the Gross-Pitaevskii 
equation~\cite{BS-2004}.

\section{Multi-D gap solitons in radially periodic potentials}
In this section we consider the dynamics of matter waves in a radial optical lattice, described by the 2D Gross-Pitaevskii equation:
\begin{equation}
iu_{t} = -\left( \frac{\partial^{2}}{\partial r^{2}} + \frac{1}{r} \frac{\partial}{\partial r} + \frac{1}{r^{2}} \frac{\partial^{2}}{\partial \theta^{2}} \right) u - \varepsilon \cos(2r) u - \chi |u|^{2} u,  \label{GPEspher}
\end{equation}
with $\chi = \pm 1$ representing attractive and repulsive nonlinearities, respectively. We seek stationary solutions of the form:
\begin{equation}
u(r,\theta,t) = \phi(r) e^{-i\mu t + i l \theta},  \label{stat}
\end{equation}
where $l$ is an integer number fixing the vorticity. Substitution into Eq.~(\ref{GPEspher}) leads to a radial equation for $\phi(r)$ which, under the transformation $\phi(r) = r^{-1/2} U(r)$, becomes:
\begin{equation}
\frac{d^{2}U}{dr^{2}} + \left[ \mu + \varepsilon \cos(2r) + \chi \frac{U^2}{r} - \frac{l^{2} + 1/4}{r^{2}} \right] U = 0. \label{reducedradial}
\end{equation}
\begin{figure}[tbh]
\centerline{
\includegraphics[width=6cm,clip]{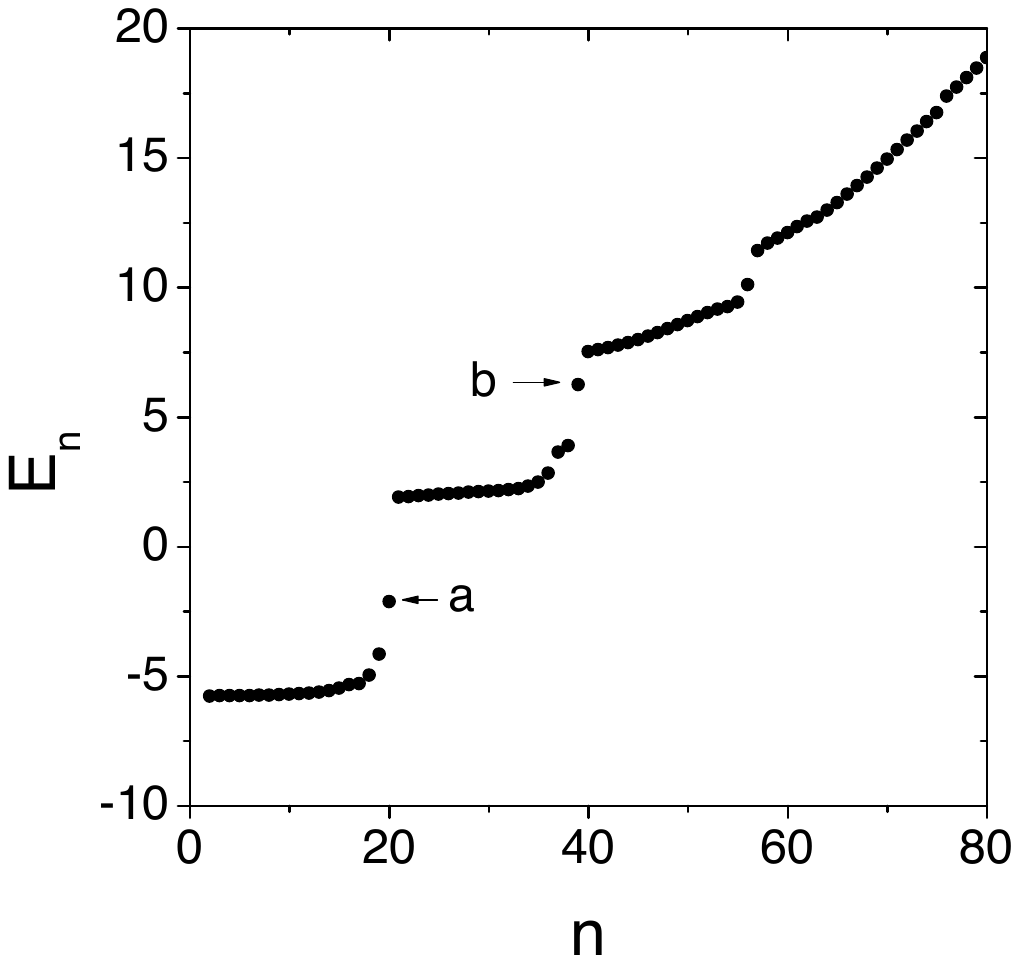}
\includegraphics[width=6cm,clip]{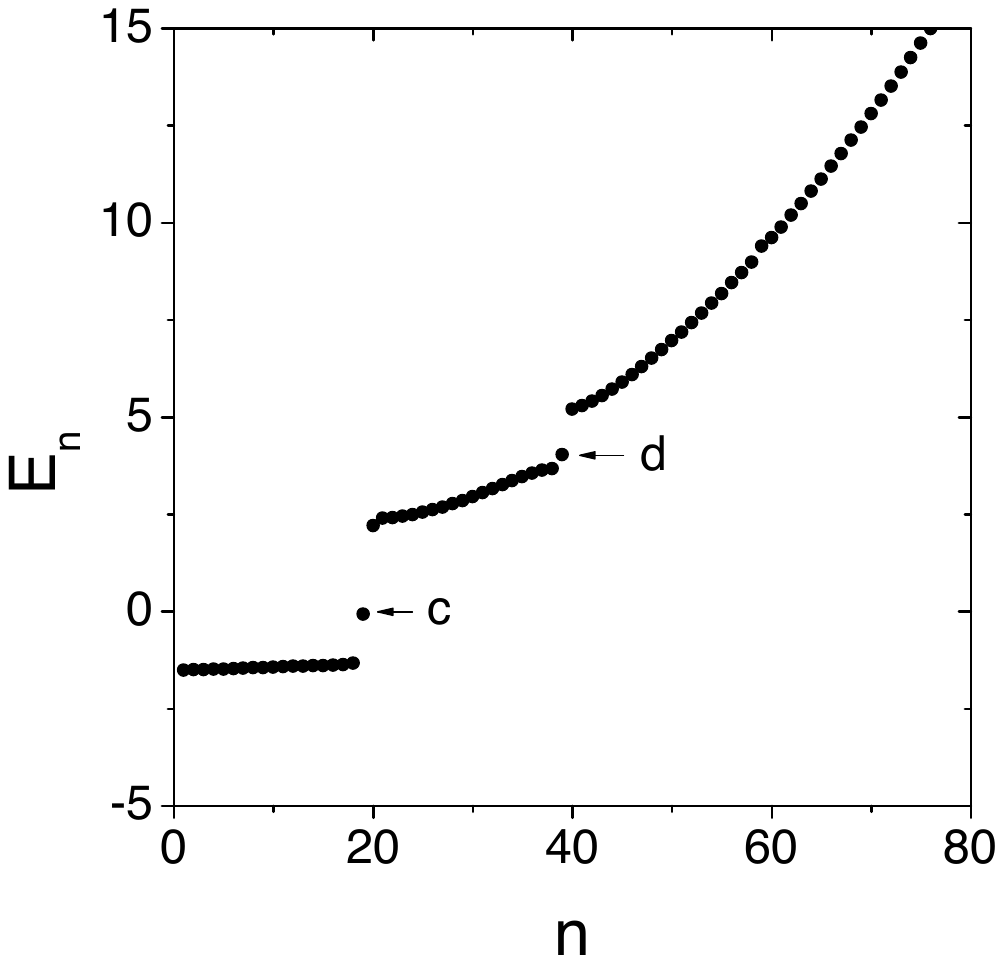}}
\vspace{-2.5cm}
\caption{Left panel: Chemical potential spectrum $(E_n \equiv \mu_n)$ obtained from the  nonlinear eigenvalue problem  (\protect\ref%
{reducedradial}) with $\protect\chi =+1$ (attraction), for $l=10$ and $%
\protect\varepsilon =-10$ ($\protect\varepsilon <0$ means the presence of a
potential maximum at $r=0$). Right panel: The same, but for the repulsive
model, $\protect\chi =-1$, with $l=2$ and $\protect\varepsilon =4$. Figure extracted from Ref.~\cite{Baizakov-2006}.}
\label{fig4}
\end{figure}
\begin{figure}[tbh]
\centerline{
\includegraphics[width=6cm,height=3cm,clip]{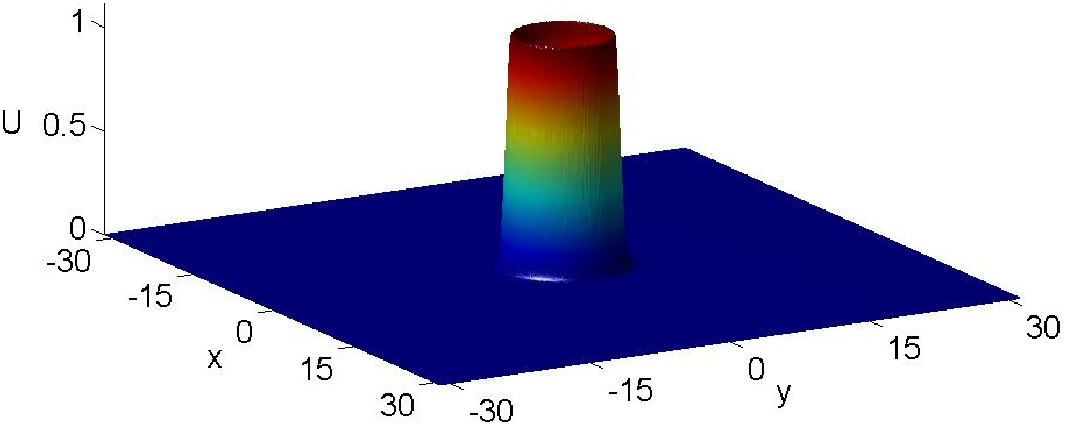}
\includegraphics[width=6cm,height=3cm,,clip]{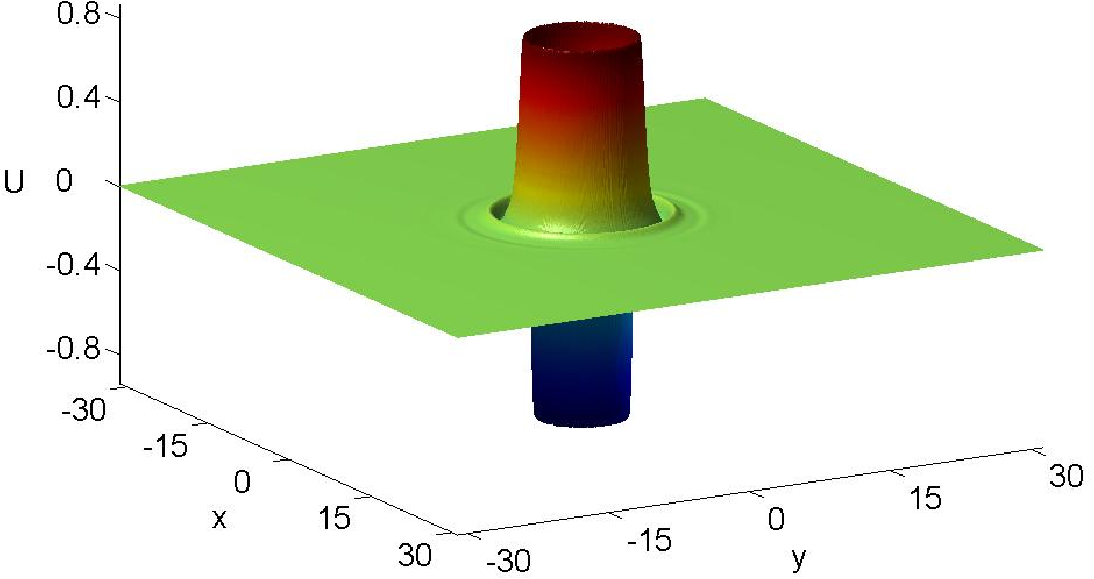}}
\centerline{
\includegraphics[width=6cm,height=3cm,clip]{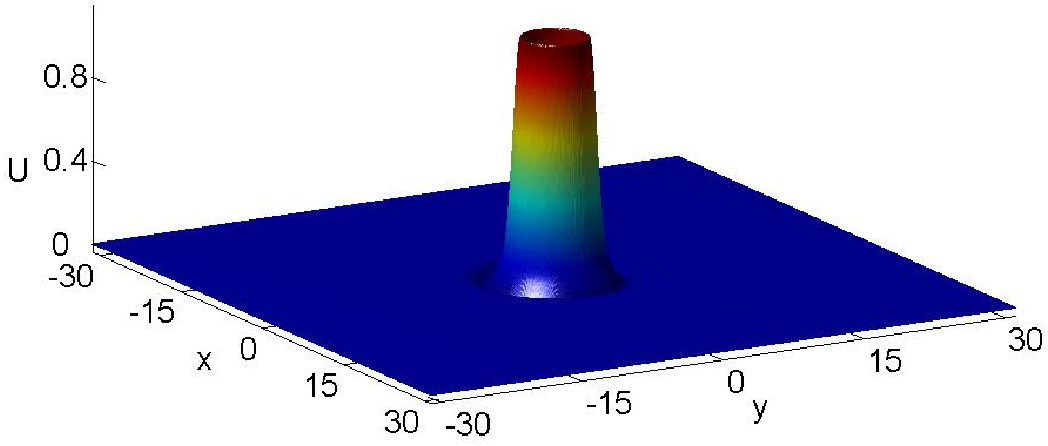}
\includegraphics[width=6cm,height=3cm,clip]{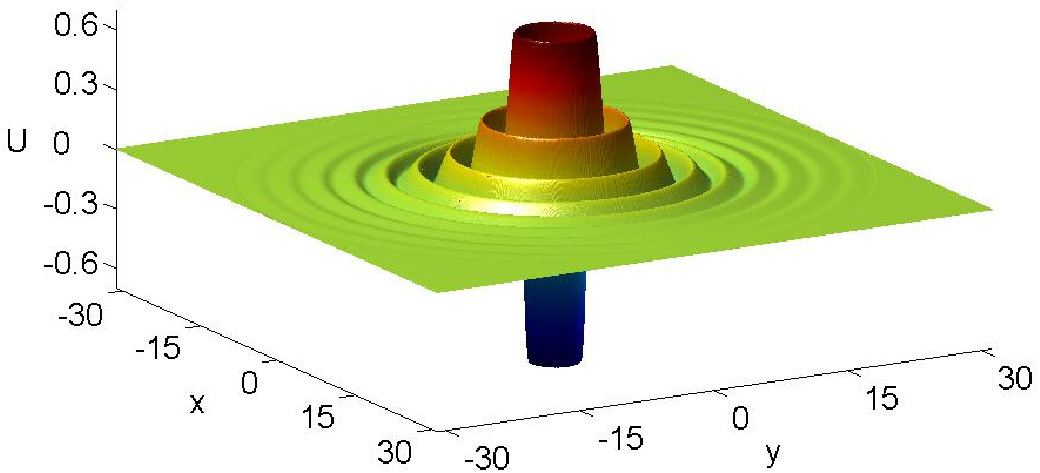}
}
\caption{Left and right top panels: 3D view of the annular gap solitons corresponding to bound states (a) and (b), respectively, of Fig.~\ref{fig4}. Left and right bottom panels: The same for bound states (c) and (d) from Fig.~\ref{fig4}.  Figure extracted from Ref.~\cite{Baizakov-2006}.}
\label{fig5}
\end{figure}
Fig.~\ref{fig4} shows the lower parts of the chemical potential  spectrum ($\mu_n\equiv E_n$)  for axisymmetric states, as found from the
effective nonlinear stationary equation, Eq.~(\ref{reducedradial}) both with 
attractive and repulsive interactions~\cite{Baizakov-2006}. Localized solutions with chemical potentials in the gaps correspond to \textit{annular gap solitons}, trapped in circular potential troughs at finite $r$. 
These spectrum and the solitons were constructed using self-consistent diagonalizations~\cite{ms05} of the nonlinear eigenvalue problem in Eq.~(\ref{stat}) and tested for stability via time-dependent simulations of Eq.~(\ref{GPEspher}). Representative 3D shapes of the soliton profiles are shown in Fig.~\ref{fig5}. They exhibit approximate onsite-symmetric or onsite-antisymmetric structures with respect to the minima of the radial potential, analogous to 1D gap solitons in OLs.

\begin{figure}[tbh]
\centerline{
\includegraphics[width=6cm,height=3cm]{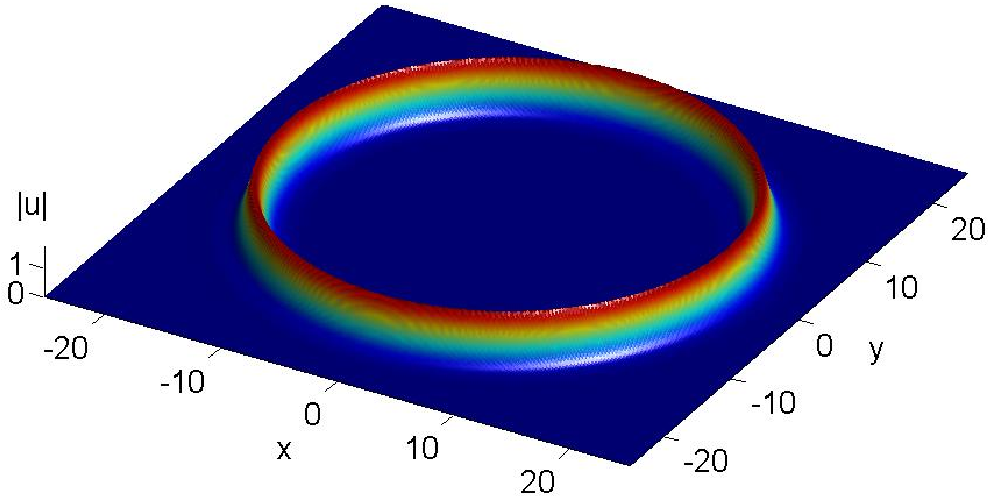}
\includegraphics[width=6cm,height=3cm]{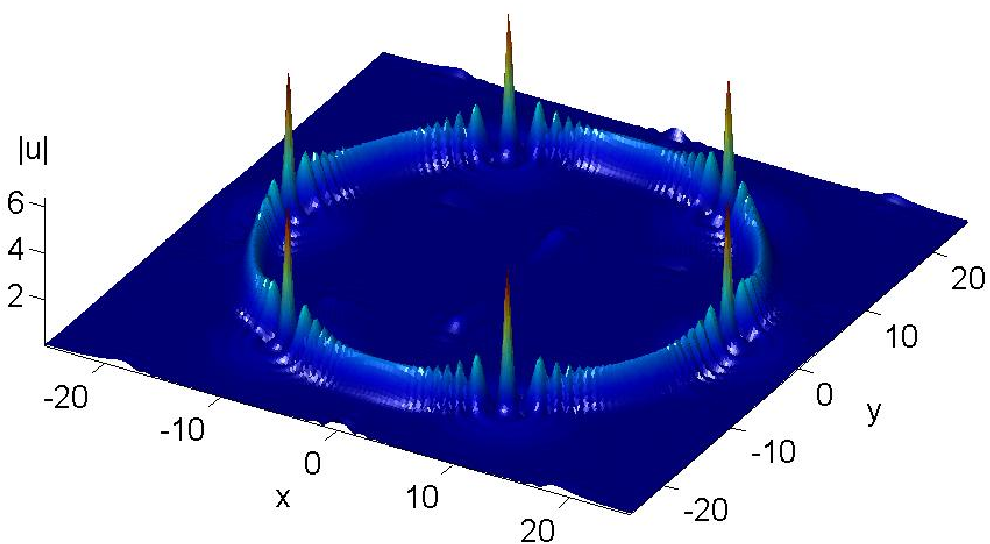}} 
\caption{(MI development in a ring soliton trapped at $r_0 = 6\pi$, with $\varepsilon=2$, amplitude 1.8, $\mu = -2.3$, and $N = 420$ (close to the predicted $N=426$). Left panel: Initial state; Right panel: necklace structure at $t=5$ generated from MI  after an azimuthal perturbation $\delta u = 0.02 \cos(6\phi) e^{-(r - r_0)^2/2}$ of the initial state. Figure adapted from Ref.~\cite{Baizakov-2006}.}
\label{fig6}
\end{figure}
\begin{figure}[tbh]
\centerline{
\includegraphics[width=6cm,height=4cm]{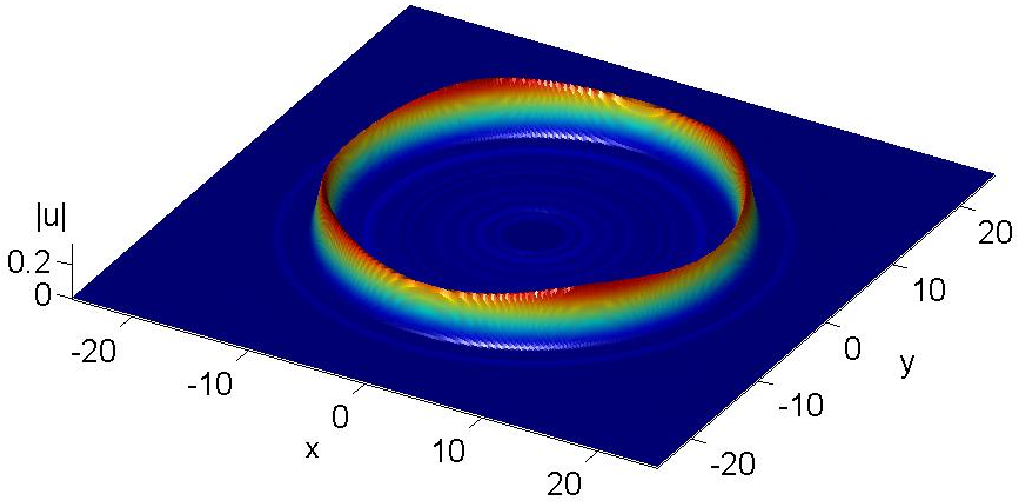}
\includegraphics[width=6cm,height=4cm]{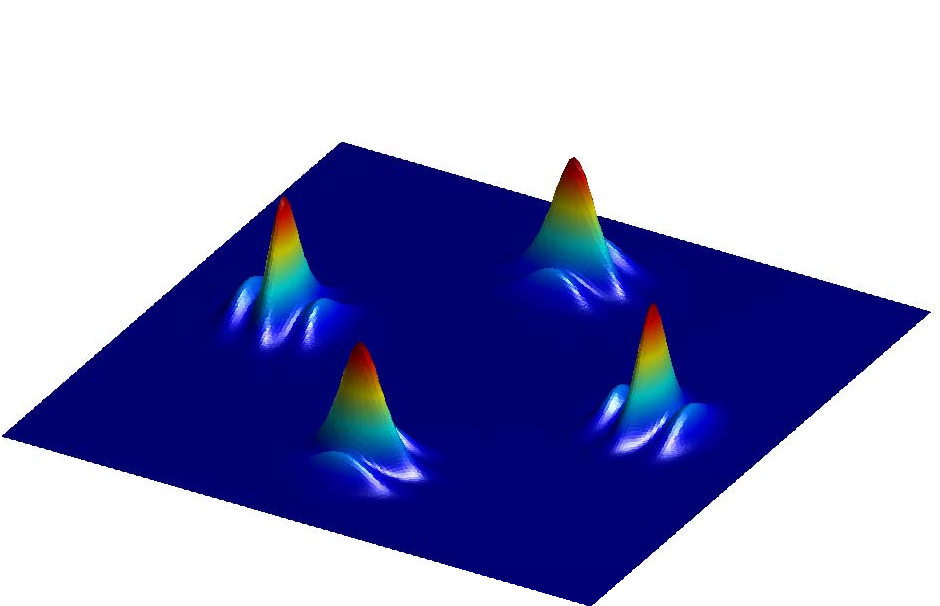}}
\caption{Left panel: A stable ring soliton with weak azimuthal modulation for $\varepsilon=10$, $N=7.2$, and $\mu = -5.7$. Right panel: Stable four-peak necklace soliton in a circular trough with $\varepsilon=2$ and $r_0 = 5\pi$, with $\pi$ phase shifts between adjacent solitons. Figure adapted from Ref.~\cite{Baizakov-2006}.}
\label{fig7}
\end{figure}

Stability analysis reveals that in the attractive case, annular solitons may be unstable due to modulational 
instability, leading to azimuthal fragmentation. Figure~\ref{fig6} illustrates a typical case, where a uniform ring-
shaped soliton becomes unstable and evolves into a necklace-like pattern composed of six localized peaks. These peaks 
can be interpreted as solitons forming on a quasi-uniform background. The initial instability is triggered by a small 
azimuthal perturbation, and the dynamics are consistent with MI analytical predictions~\cite{Baizakov-2006}.

Azimuthally modulated stationary rings corresponding to cnoidal wave solutions were also observed, as shown in the 
left panel of Fig.~\ref{fig7}. This state emerges from MI of uniform ring soliton and transitions to necklace states 
with increasing norm. Stronger modulation corresponds to patterns as in the right panel of Fig.~\ref{fig6}, while weaker 
modulation yields smoother structures like the one in the left panel of  Fig.~\ref{fig7}.

The right panel of Figure~\ref{fig7} shows a stable configuration of four solitons, equally spaced in the same trough of the potential, obtained not from MI but from 
imaginary-time evolution of four Gaussian pulses with alternating phases. Each localized soliton has norm $N = 2\pi$, 
and the structure remains stable due to the radial trapping and phase arrangement~\cite{Baizakov-2006}.

Strongly localized solitons in the potential troughs can undergo persistent circular motion if given a small initial 
velocity. At higher velocities, centrifugal effects induce leakage into adjacent radial channels. The soliton's norm 
decreases due to tunneling losses, leading to its eventual disintegration when the norm drops below a critical 
threshold~\cite{Baizakov-2006}. If the radial lattice is shallow and the soliton is strongly self-trapped, the centrifugal force can push the entire soliton outward, eventually leading to decay through radiation losses. 

Collisions between solitons of the type shown in the right panel of Fig.~\ref{fig7}, traveling in the same or in adjacent 
circular troughs, were also investigated. Results from GPE numerical integrations show that while collisions between in-phase solitons in a common trough lead to collapse, out-of-phase solitons may bounce many times and eventually merge into a single one, without collapsing. In-phase solitons colliding in adjacent circular troughs also tend to merge into a single soliton~\cite{Baizakov-2006}.

\subsection{3D solitons in cross-combined linear and nonlinear OLs}
Beyond standard linear optical lattices (LOLs), multidimensional solitons can also be stabilized using nonlinear 
optical lattices (NOLs), created via spatially modulated interactions—e.g., by tuning Feshbach resonances using 
magnetic or optical fields~\cite{OFR2}. In particular, 3D solitons can be stabilized using cross-combined LOL and NOL 
configurations~\cite{Abdullaev-2012,Ismailov-2024}.

A model for a BEC in such a setting is described by the normalized GP equation:
\begin{equation}
{\rm i} \hbar \frac{\partial \Psi}{\partial t}= - \frac{\hbar^2}{2 m} \nabla^2 \Psi
- v_0 [\cos(2 k\,x) +  \cos(2 k\,z)] \Psi - g(y)|\Psi|^2 \Psi,
\label{3DGPEX}
\end{equation}
where $g(y)= g_0 + g_1 \cos(2 \kappa  y)$ modulates the interaction in the $y$-direction.
\begin{figure}[tbp]
\centerline{
\includegraphics[width=12.5cm,clip]{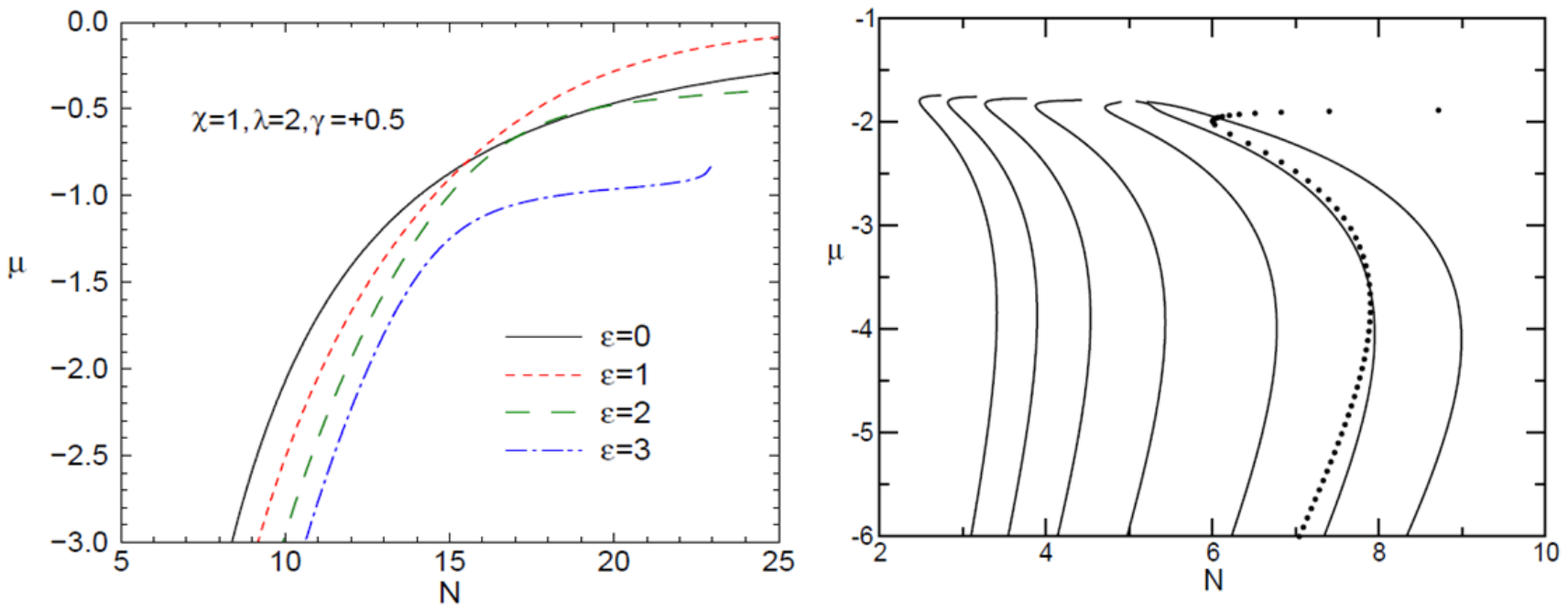}
}
\vskip -1.5cm
\caption{Left panel: Chemical potential $\mu$ versus atom number $N$ obtained from the variational approximation (VA) for 3D solitons in a system with crossed 1D linear (LOL, along $x$) and nonlinear (NOL, along $y$) optical lattices, without confinement along $z$ ($\varepsilon_z = 0$) and attractive interactions ($\chi = 1$). Other parameters are given in the frame.  
Right panel: Same as in left panel, but with a 2D LOL in the $x$ and $z$ directions and a 1D NOL along $y$. Parameters: $\varepsilon = 3$, $\chi = 1$, with NOL strength $\gamma$ varying from 3.5 to 0.5 (curves from left to right). Dotted line shows PDE results for $\gamma = 0.5$ for comparison. Figure adapted from Ref.~\cite{Abdullaev-2012}.
}
\label{fig8}
\end{figure}

Introducing dimensionless variables via $t \rightarrow (\hbar/E_r)\,t$, ${\mathbf r} \rightarrow {\mathbf r}/k$, and 
$\Psi \rightarrow k^{3/2} \psi$, with $E_r = \hbar^2 k^2 / 2m$, the equation becomes:
\begin{equation}
{\rm i}\frac{\partial \psi}{\partial t} = - \nabla^2 \psi - V_{l} \psi - V_{nl} |\psi|^2 \psi,
\end{equation}
where
\beq
V_l = \varepsilon (\cos(2x) + \cos(2z)), \quad
V_{nl} = \chi + \gamma \cos(\lambda y),
\eeq
with $\varepsilon = v_0/E_r$, $\chi = 8 \pi a_0 k$, $\gamma = g_1 k^3/E_r$, and $\lambda = 2\kappa/k$.
Assuming a Gaussian ansatz for stationary solitons:
\begin{equation}
U(x,y,z)= A\, e^{- \frac{1}{2} (a x^2 + b y^2 + c z^2)},
\end{equation}
and computing the effective Lagrangian yields:
\begin{equation}
L_{eff} = \frac{\pi^{3/2} A^2}{4 a \sqrt{b}}\left[2 a + b  -2(\mu + 2 \varepsilon e^{-1/a})
- \frac{A^2}{2\sqrt{2}}(\chi + \gamma e^{-\lambda^2/8b})\right],
\end{equation}
with $N = A^2 \pi^{3/2}/(a \sqrt{b})$.
Minimizing $L_{eff}$ leads to:
\begin{equation}
\mu = \frac{b}{2} -a -2\varepsilon\left(1-\frac{2}{a}\right) e^{-1/a}, \quad
N = 4\pi\sqrt{\frac{2\pi}{b}}\,\frac{1 - 2\varepsilon e^{-1/a}/a^2}{\chi + \gamma  e^{-\lambda^2/8b}},
\end{equation}
with $a$ and $b$ related by:
\begin{equation}
\frac{1}{b} + \frac{\lambda^2 \gamma e^{-\lambda^2/8b}}{4 b^2 (\chi + \gamma e^{-\lambda^2/8b})}
= \frac{1}{a - (2\varepsilon/a) e^{-1/a}}.
\label{trans2}
\end{equation}

Fig.~\ref{fig8} shows VA curves in the $(\mu, N)$ plane, illustrating the existence of stable 3D solitons  in a cross-
combination of 2D linear OL in the $x$ and  $z$ directions and a 1D NOL in the y-direction (see right panel). 
\begin{figure}
\vspace{-1cm}
\centerline{
\includegraphics[width=12.5cm,clip]{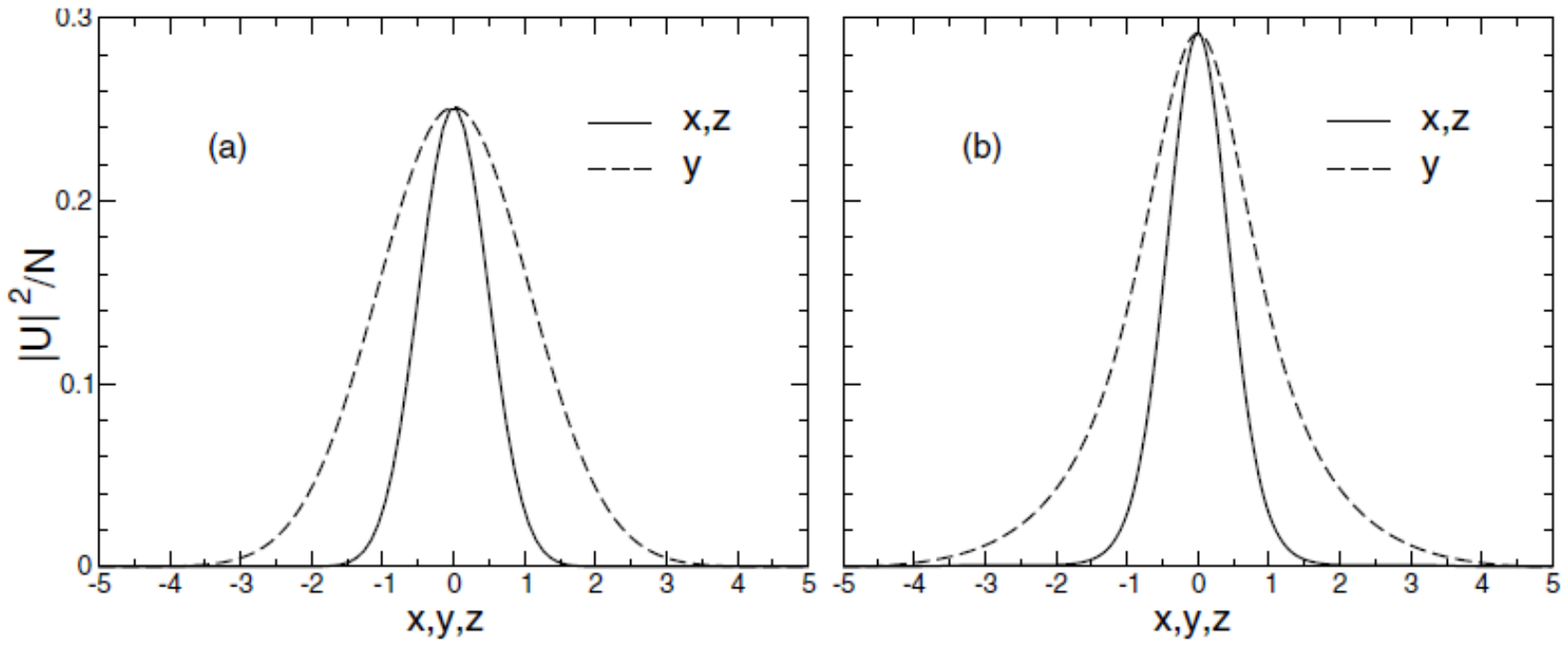}
}
\vskip -1.6cm
\caption{Soliton profile sections from VA (left) and PDE (right) simulations for $\chi = 1$, $\gamma = 0.5$. Agreement between both approaches is evident. Figure extracted from Ref.~\cite{Abdullaev-2012}.
}
\label{fig9}
\end{figure}
From the left panel, we observe that no stable 3D soliton forms in the cross-combined configuration of a 1D linear optical lattice (LOL) along $x$ and a nonlinear optical lattice (NOL) along $y$, with no confinement in the $z$-direction. 

In both configurations, soliton stability can be predicted by the Vakhitov-Kolokolov (VK) criterion and confirmed through numerical simulations of the Gross-Pitaevskii equation (GPE). These results indicate that NOLs are generally less effective than LOLs in stabilizing multidimensional solitons.

The comparison between VA predictions and full GPE simulations, shown in Fig.~\ref{fig8} (dotted lines) and Fig.~\ref{fig9} (soliton profiles), reveals good agreement in both cases.

\section{Stabilization of Multi-D Solitons via Parameter Management}

The collapse of multidimensional solitons in Bose-Einstein condensates (BECs) with attractive interactions has prompted the development of external control strategies. Among the most effective are dynamical modulation techniques, which stabilize solitons by time-dependent tuning of system parameters.
In nonlinear management, the interaction strength is periodically modulated, typically via Feshbach resonances, effectively averaging the nonlinearity to suppress collapse~\cite{Abdullaev-2003, Saito-2003}. Similarly, Rabi management employs coherent coupling between atomic states, induced by external fields, to redistribute populations and reshape the effective interaction landscape~\cite{Abdullaev-2008, Saito-2007}.
\subsection{Nonlinear management of multi-D solitons}
A powerful method to stabilize multidimensional solitons against collapse is through rapid, periodic modulation of the
atomic scattering length. Such temporal variations can be experimentally realized using Feshbach resonance techniques.
The atomic scattering length under the influence of an external magnetic field is given by
\begin{equation}
a_s = a_{\mathrm{bg}} \left(1 - \frac{\Delta}{B(t) - B_0} \right),
\end{equation}
where $B_0$ is the resonance position, $\Delta$ is its width, and $a_{\mathrm{bg}}$ is the background scattering 
length. By modulating the magnetic field $B(t)$ in time, one induces a periodic variation of $a_s(t)$, and hence of 
the nonlinear interaction strength $\gamma(t)$.

The GPE with time-modulated two-body interaction takes the form
\begin{equation} \label{eq:LD}
i u_t + \nabla^2 u + \gamma(t) |u|^2 u = 0,
\end{equation}
where the time-dependent nonlinearity is modeled as $\gamma(t) = \gamma_0 + \gamma_1(t)$. As demonstrated in 
\cite{Abdullaev-2003, Saito-2003, Montesinos}, dynamically stabilized 2D solitons can exist. The key arguments are 
based on analyzing the averaged dynamics of the soliton width under the rapid modulation of the scattering length.
In the regime of strong management, where $\gamma_1 \sim 1/\epsilon$ and the modulation frequency is $\sim 1/\epsilon$ 
with $\epsilon \ll 1$, Eq.~\eqref{eq:LD} can be averaged using the transformation~\cite{ZP, KS}
\begin{equation}
u = v \exp\left( i \Gamma(t) |v|^2 \right), \quad \Gamma(t) = \int_0^{\tau} \gamma_1(\tau')\, d\tau' - \int_0^1 
\int_0^{\tau} \gamma_1(\tau')\, d\tau'\, d\tau.
\end{equation}
This transformation removes the rapidly varying terms from the equation. Expanding $v = w + \epsilon w_1 + \epsilon^2 
w_2 + \cdots$ and averaging over fast oscillations leads to the effective GP equation:
\begin{equation}
i w_t + \nabla^2 w + \gamma_0 |w|^2 w + \sigma^2 \left( |\nabla |w|^2|^2 + 2 |w|^2 \Delta |w|^2 \right) w = 0.
\end{equation}
The correction to the energy due to the modulated nonlinearity is
\begin{equation}
E_{\mathrm{mn}} = \sigma^2 (\nabla |w|^2)^2 |w|^2.
\end{equation}
This expression is also valid in the \textit{weak management} regime ($\sigma^2 \ll 1$)~\cite{AMKT}.

A scaling analysis yields the effective energy of the system as:
\begin{equation} \label{eq:ener}
\langle E \rangle = \frac{A}{L^2} - \frac{B}{L^D} + \frac{C}{L^{2D + 2}}, \quad C \sim \sigma^2,
\end{equation}
where $L$ is the characteristic soliton width and $D$ is the spatial dimensionality. This energy functional possesses a minimum, indicating the existence of stable 2D and 3D solitons under nonlinearity management.
\begin{figure}[ht]
\centerline{
\includegraphics[width=14cm]{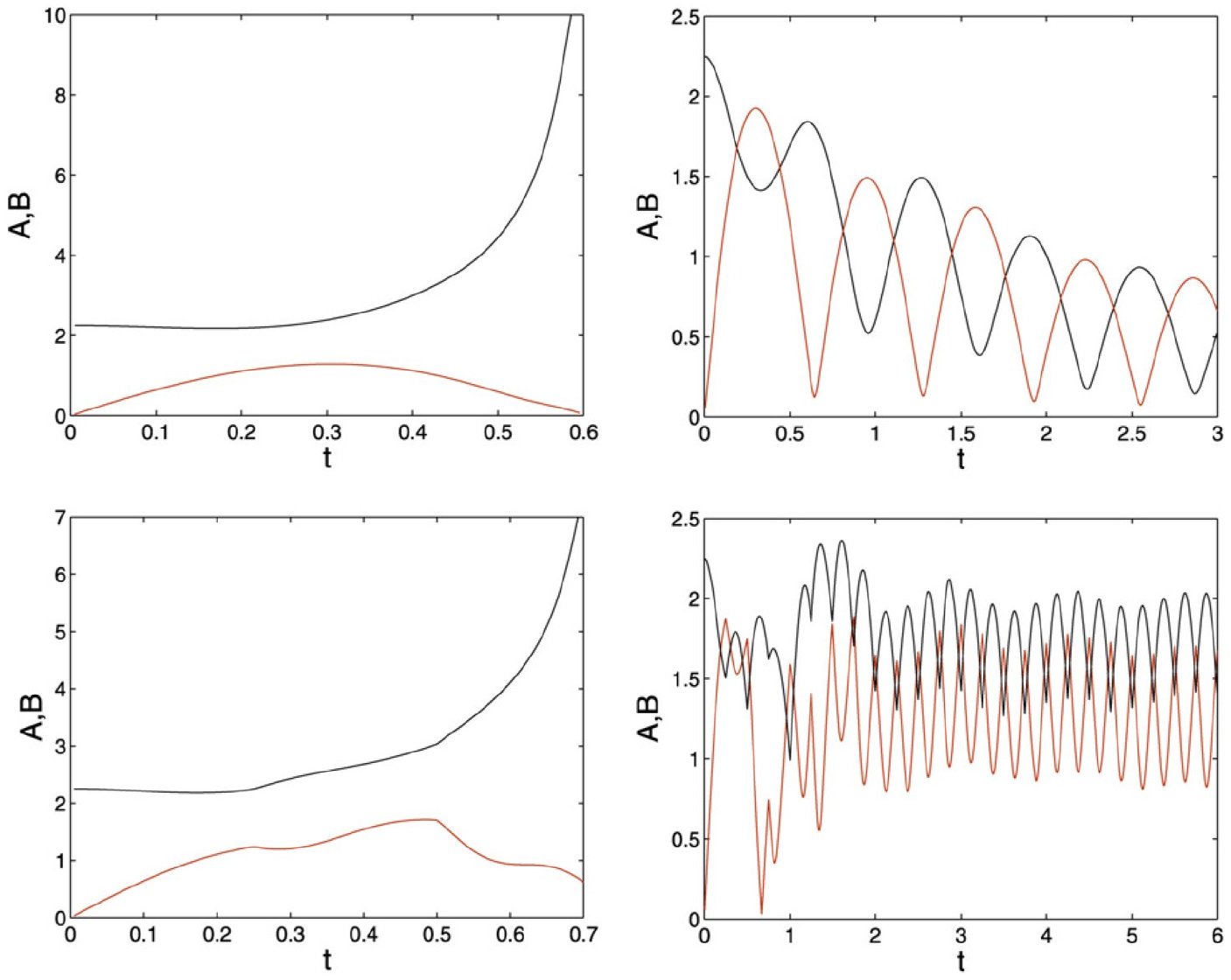}}
\caption{
Time evolution of the component amplitudes $\{A(t), B(t)\}$ from simulations of the full GP system. Top: dynamics for 
constant Rabi frequencies $\Omega/B = 3$ (left) and 5 (right). Bottom: results for time-modulated $\Omega(t)$ as given 
by Eq.~(\ref{eq:mod1}), with period $T = 1/2$. Figure extracted from Ref.~\cite{Abdullaev-2008}.}
\label{fig10}
\end{figure}
The same result can be derived using a modified variational approach that incorporates rapid self- and cross-phase modulations induced by strong nonlinearity management~\cite{AG}. The variational ansatz for the wavefunction is
\begin{equation} \label{eq:ansatz1}
u = A Q\left( \frac{\rho}{w} \right) \exp\left( -\frac{\rho^2}{2w^2} + i b \rho^2 + i \phi \right) \exp\left[ i \Gamma\left( \frac{t}{\epsilon} \right) Q^2\left( \frac{\rho}{w} \right) A^2 \right],
\end{equation}
where
\begin{equation}
Q\left( \frac{\rho}{w} \right) = \exp\left( -\frac{\rho^2}{2w^2} \right), \quad \rho^2 = x^2 + y^2,
\end{equation}
and $\Gamma$ is the zero-mean antiderivative of $\gamma_1(t)$.
Substituting the ansatz into Eq.~\eqref{eq:LD} and averaging over the rapid modulations yields the same expression for the averaged energy as in Eq.~\eqref{eq:ener}, providing further justification for the validity of the averaging method. Some arguments for a slow decay of nonlinearity-managed solitons at large propagation times have been presented in~\cite{Itin}. The stabilization of 2D vector solitons via numerical simulations has been explored in~\cite{Montesinos-2004}, and the general theory of nonlinearity management for vector solitons was developed in~\cite{Abdullaev-2023}.
\subsection{Stabilization via Rabi management}
The possibility of stabilizing 2D solitons via time-periodic modulation of the atomic scattering length opens the door 
to other effective management schemes. One such mechanism involves exploiting the Rabi coupling between components of 
a two-component Bose–Einstein condensate (BEC). 

An external electromagnetic field, resonantly coupling two hyperfine states, induces periodic transitions of atoms 
between these states. This leads to time-dependent population imbalances, effectively modulating the mean-field 
interactions, including changes in sign from repulsive to attractive and vice versa. This mechanism provides a pathway 
for the dynamical stabilization of 2D solitons. Below, we discuss the two cases of time constant and time-dependent Rabi frequency. 

When the Rabi frequency $\Omega$ is constant, the two-component GP system can be transformed via the rotation:
\begin{equation}
U = u\sin(\Omega t) + v\cos(\Omega t), \quad V = u\cos(\Omega t) + v\sin(\Omega t).
\end{equation}
Assuming $v(t=0)=0$, then $V \approx 1/\Omega \ll 1$ and the system reduces to a single NLSE with a time-periodic nonlinearity\cite{Saito-2007}:
\begin{equation}
i U_t = -\frac{1}{2} \nabla^2 U + \gamma(t)|U|^2 U, \quad \gamma(t) = 
\gamma_0 + \gamma_1 \cos(2\Omega t) + \gamma_2 \cos(4\Omega t),
\end{equation}
where 
$$
\gamma_0=(3(g_{11}+g_{22})+2g_{12})/8,\ \gamma_1 =(g_{11}-g_{22})/2,\ \gamma_=(g_{11}+g_{22}+2g_{12})/8.
$$
An analysis analogous to that of nonlinear management (see previous section) demonstrates that stabilization is 
achievable for sufficiently large $\Omega$. The variational analysis yields the characteristic soliton size:
\begin{equation}
R_c \approx \left( \frac{3N^2 (4\gamma_1^2 + \gamma_2^2)}{4\pi \Omega^2
+ N \gamma_0} \right)^{1/4}.
\end{equation}
This theoretical prediction aligns well with results from direct numerical simulations of the full coupled GP system.

In the second case, i.e. when $\Omega(t)$ is rapidly and periodically modulated, one can derive an 
averaged system of coupled GP equations\cite{Abdullaev-2008}. This system includes nonlinear tunneling terms of the form $\phi_0^2 \psi_0$ and 
$\psi_0^2 \phi_0$, where $\psi_0,\phi_0$ are zero-order terms of the averaged expansions for fields, and represent higher-order effective interactions between the components.

Numerical simulations of both the original and the averaged systems show that long-lived 2D solitons can be sustained 
under this type of Rabi management. For instance, consider the modulation:
\begin{equation}\label{eq:mod1}
\frac{1}{\epsilon} \Omega\left( \frac{t}{\epsilon} \right) = 
\begin{cases}
M, & \text{if } 0 < \operatorname{mod}(\tau, T) < T/2, \\
-M, & \text{if } T/2 < \operatorname{mod}(\tau, T) < T,
\end{cases}
\quad \text{with } \tau = \frac{t}{\epsilon}, \quad T = \frac{1}{2}.
\end{equation}

These results indicate that when the linear coupling coefficient is periodically modulated, 2D solitons can remain 
localized for extended times even in free space- {see Fig.~\ref{fig10}. Moreover, if a weak external trapping potential is added, the soliton may become completely stabilized.

\section{Conclusions}
The study of attractive multidimensional solitons in the nonlinear Schrödinger equation (NLS) family is a rich field 
at the intersection of theory, simulations, and experiments. Theoretically, the multidimensional NLSE (or GPE) has 
been a key framework for understanding collapse dynamics, modulational instability, and soliton 
formation~\cite{Sulem-1999, Berge}. Numerical simulations have confirmed analytical predictions, clarified scenarios
like collapse arrest, and identified stability regimes~\cite{Fibich-1999, Pitaevskii-2003}. Experimentally, matter-
wave soliton trains~\cite{Strecker-2002, Cornish-2006}, Bosenova-like collapses~\cite{Donley-2001}, and soliton 
dynamics in optical lattices~\cite{Becker-2008} have validated many theoretical predictions while revealing complex 
effects beyond the mean-field approximation.

Despite progress, several key challenges remain. While mechanisms for stabilizing 2D solitons have been proposed, 
achieving robust stabilization of fully 3D solitons in attractive BECs or photonic systems is still elusive. Quantum 
fluctuations, thermal effects, and beyond-mean-field corrections become crucial near collapse and in low-atom-number 
regimes~\cite{Petrov-2015, Wild-2012}, but their interaction with modulational instability and soliton formation 
remains poorly understood.

In photonics, engineered 2D and 3D solitons offer potential for advanced light localization and switching 
schemes~\cite{Kartashov-2011, Malomed-2023b}, though practical implementation is still in early stages. The inclusion 
of synthetic gauge fields and spin-orbit coupling introduces further complexity and opportunities in both BECs and 
nonlinear optics~\cite{Lin-2011, Achilleos-2013}, opening new directions for studying multidimensional soliton 
stability, dynamics, and interactions.
\vskip .3cm
\noindent {\bf Acknowledgments} The authors wish to thank Prof. P.G. Kevrekidis for interesting discussions and suggestions.
\vskip .5cm
\noindent This preprint is the version prior to its acceptance as a chapter of the book {\it Short and Long Range Quantum Atomic Platforms – Theoretical and Experimental Developments} (provisional title) edited by P.G. 
Kevrekidis , C.L. Hung, and S. I. Mistakidis, Springer Tracts in Modern Physics Series,  Springer Nature Switzerland AG, Gewerbestrasse 11, 6330 Cham, Switzerland.

\end{document}